\documentclass[letterpaper, 10pt, conference]{ieeeconf}
\let\proof\relax

\usepackage{algorithm}
\usepackage{algpseudocode}
\usepackage{pifont}
\usepackage{multirow}
\usepackage[T1]{fontenc}
\usepackage{amsthm,xpatch}
\usepackage{amsmath,amsfonts}
\usepackage{cite}
\usepackage{amssymb}
\usepackage{dsfont}
\usepackage{color}
\usepackage{breqn}
\usepackage{mathtools}
\usepackage{bbm}
\usepackage{latexsym}
\usepackage{accents}
\usepackage{tikz}
\usepackage[mathscr]{eucal}
\usepackage{caption}
\usepackage{subcaption}
\usepackage{booktabs}
\usetikzlibrary{calc}
\usetikzlibrary{math}
\usetikzlibrary{arrows.meta}
\usepackage{csquotes}
\usepackage{varioref}

\DeclareMathOperator*{\argmax}{\arg\!\max}

\newcommand{\abs}[1]{\lvert{#1}\rvert}

\newcommand*{\transpose}{%
  {\mathpalette\@transpose{}}%
}

\IEEEoverridecommandlockouts

\begin{document}

\newcommand{\SB}[3]{
\sum_{#2 \in #1}\biggl|\overline{X}_{#2}\biggr| #3
\biggl|\bigcap_{#2 \notin #1}\overline{X}_{#2}\biggr|
}

\newcommand{\Mod}[1]{\ (\textup{mod}\ #1)}

\newcommand{\overbar}[1]{\mkern 0mu\overline{\mkern-0mu#1\mkern-8.5mu}\mkern 6mu}

\makeatletter
\newcommand*\nss[3]{%
  \begingroup
  \setbox0\hbox{$\m@th\scriptstyle\cramped{#2}$}%
  \setbox2\hbox{$\m@th\scriptstyle#3$}%
  \dimen@=\fontdimen8\textfont3
  \multiply\dimen@ by 4             % 4x the default rule thickness
  \advance \dimen@ by \ht0
  \advance \dimen@ by -\fontdimen17\textfont2
  \@tempdima=\fontdimen5\textfont2  % x-height
  \multiply\@tempdima by 4
  \divide  \@tempdima by 5          % 80% of the x-height
  % Modifications are only necessary if the top of the subscript is not that high:
  \ifdim\dimen@<\@tempdima
    \ht0=0pt                        % don't let the subscript interfere
    \@tempdima=\fontdimen5\textfont2
    \divide\@tempdima by 4          % 25% of the x-height
    \advance \dimen@ by -\@tempdima % if >0, add to depth of superscript!
    \ifdim\dimen@>0pt
      \@tempdima=\dp2
      \advance\@tempdima by \dimen@
      \dp2=\@tempdima
    \fi
  \fi
  #1_{\box0}^{\box2}%
  \endgroup
  }
\makeatother

\makeatletter
\renewenvironment{proof}[1][\proofname]{\par
  \pushQED{\qed}%
  \normalfont \topsep6\p@\@plus6\p@\relax
  \trivlist
  \item[\hskip\labelsep
        \itshape
%    #1\@addpunct{.}]\ignorespaces% DELETED
    #1\@addpunct{:}]\ignorespaces% ADDED
}{%
  \popQED\endtrivlist\@endpefalse
}
\makeatother

\makeatletter
\newsavebox\myboxA
\newsavebox\myboxB
\newlength\mylenA

\newcommand*\xoverline[2][0.75]{%
    \sbox{\myboxA}{$\m@th#2$}%
    \setbox\myboxB\null% Phantom box
    \ht\myboxB=\ht\myboxA%
    \dp\myboxB=\dp\myboxA%
    \wd\myboxB=#1\wd\myboxA% Scale phantom
    \sbox\myboxB{$\m@th\overline{\copy\myboxB}$}%  Overlined phantom
    \setlength\mylenA{\the\wd\myboxA}%   calc width diff
    \addtolength\mylenA{-\the\wd\myboxB}%
    \ifdim\wd\myboxB<\wd\myboxA%
       \rlap{\hskip 0.5\mylenA\usebox\myboxB}{\usebox\myboxA}%
    \else
        \hskip -0.5\mylenA\rlap{\usebox\myboxA}{\hskip 0.5\mylenA\usebox\myboxB}%
    \fi}
\makeatother

\xpatchcmd{\proof}{\hskip\labelsep}{\hskip3.75\labelsep}{}{}

\pagestyle{plain}

\title{\fontsize{22.59}{28}\selectfont 
%The Effect of Scheduling in Belief Propagation Decoding for Noisy Group Testing
Scheduling Improves the Performance of Belief Propagation for Noisy Group Testing
}

\author{Esmaeil Karimi, Anoosheh Heidarzadeh, Krishna R. Narayanan, and Alex Sprintson\thanks{The authors are with the Department of Electrical and Computer Engineering, Texas A\&M University, College Station, TX 77843 USA (E-mail: \{esmaeil.karimi, anoosheh, krn, spalex\}@tamu.edu).}}%\thanks{This material is based upon work supported by the National Science Foundation under Grants No. 1718658, 1642983, and 1547447.}}

%\thanks{This work was supported by the National Science Foundation under Grant No.~CNS-0954153 and the AFOSR under Contract No.~FA9550-13-1-0008.}

\maketitle 

\thispagestyle{plain}

\begin{abstract}
This paper considers the noisy group testing problem where among a large population of items some are defective. The goal is to identify all defective items by testing groups of items, with the minimum possible number of tests. 
The focus of this work is on the practical settings with a limited number of items rather than the asymptotic regime. 
In the current literature, belief propagation 
has been shown to be effective in recovering defective items from the test results.
In this work, we adopt two variants of the belief propagation algorithm for the noisy group testing problem. 
These algorithms have been used successfully in decoding of low-density parity-check codes. 
We perform an experimental study and using extensive simulations we show that these algorithms 
achieve higher success probability, lower false-negative and false-positive rates compared to the traditional belief propagation algorithm.
For instance, our results show that the proposed algorithms can reduce the false-negative rate by about $50\%$ (or more) when compared to the traditional BP algorithm, under the combinatorial model. Moreover, under the probabilistic model, this reduction in false-negative rate increases to about $80\%$ for the tested cases.

\end{abstract}

\section{introduction}
We consider the noisy Group Testing (GT) problem which is concerned with recovering all defective items in a given population of items. In the group testing problem, the result of a test on any group of items is binary. The objective is to design a test plan for the group testing problem with a minimum number of tests. Aside from the theoretical endeavors, the GT problem has also gained substantial attention from the practical perspective. In particular, the GT problem has been studied for a wide range of applications from biology and medicine\cite{7805159} to information and communication technology\cite{6739161,8891708}, and computer science\cite{Malioutov2017}. % over the last few years 
Very recently, group testing has also been used for COVID-19 detection~\cite{DBLP:journals/corr/abs-2004-04785,10.1093/ajcp/aqaa064,aldridge2020conservative,Shentaleabc5961}.

There are two different scenarios for the defective items. In the combinatorial model, the exact number of defective items is known, whereas in the probabilistic model, each item is defective with some probability, independent of the other items \cite{8989403,8635972,DBLP:journals/corr/abs-2007-08111}. In this work, we consider both models. In the combinatorial model, we assume that there are exactly $K$ defective items among a population of $N$ items. In the probabilistic model, we suppose that each item is defective with probability $\frac{K}{N}$, independently from the other items, where $N$ is the total number of items, and the parameter $K$ represents the expected number of defective items.

In this paper, we are interested in \emph{non-adaptive} group testing schemes, where all tests are designed in advance. This is in contrast to \emph{adaptive} schemes, in which the design of each test depends on the results of the previous tests~\cite{8437774,8919896,DBLP:journals/corr/abs-2101-02405}. In most practical applications, when compared to adaptive group testing schemes, non-adaptive schemes are preferred because all tests can be executed at once in parallel. Different decoding algorithms such as linear programming, combinatorial orthogonal matching pursuit, definite defectives, belief propagation (BP), and separate decoding of items have been proposed for noisy non-adaptive group testing. A thorough review and comparison of these algorithms is provided in the survey paper\cite{CIT-099}. 
It is difficult to analyze the performance of these algorithms in the non-asymptotic regime.
However, empirical evidence suggests that the BP algorithm results in lower error probabilities compared to other algorithms.

BP decoding is an iterative algorithm that passes messages over the edges in the underlying 
factor graph according to a schedule.
For a cycle-free factor graph, BP decoding is equivalent to maximum-likelihood decoding. 
However, in the presence of loops in the factor graph, BP becomes suboptimal.
The most popular scheduling strategy in BP decoding is flooding, or simultaneous scheduling, where in every iteration all the variable nodes are updated simultaneously using the same pre-update information, followed by updating all the test nodes of the graph, again, using the same pre-update information. Several studies have investigated the effects of different types of sequential, or non-simultaneous, scheduling strategies in BP for decoding low-density parity-check (LDPC) codes among which are random scheduling BP and node-wise residual BP (see~\cite{4288829} and references therein). It has been shown that sequential BP algorithms converge faster than traditional BP. %This increase in convergence speed is achieved at no cost since sequential updating does not increase the decoding complexity per iteration. 
Also, sequential updating solves some standard trapping set errors\cite{1599818,4288829}. To the best of our knowledge, these algorithms have not been used in the context of group testing.

\subsection{Main Contributions}
In this work, we focus on a practical regime in which the number of items is in the order of hundreds, and investigate the performance of two variants of BP algorithm for decoding of noisy non-adaptive group testing under %the combinatorial and probabilistic
two different models (combinatorial and probabilistic) for defective items. 
Through extensive simulations, we show that the proposed algorithms achieve higher success probability and lower false-negative and false-positive rates when compared to the traditional BP algorithm. 
%outperform the BP algorithm for all problem parameters being considered, in the sense of success probability, false-negative rate and false-positive rate. 
For instance, our results show that the proposed algorithms can reduce the false-negative rate by about $50\%$ (or more) when compared to the traditional BP algorithm, under the combinatorial model. Moreover, under the probabilistic model, this reduction in false-negative rate increases to about $80\%$ for the tested cases. 
%In addition, we show that there is a trade-off between the false-negative rate and the false-positive rate under the probabilistic model. Depending on the application at hand, one can choose a threshold value that that achieves the desired false-negative rate or false positive rate. 

\section{Problem Setup and Notations}\label{sec:SN}
Throughout the paper, we denote vectors and matrices by bold-face small and capital letters, respectively. For an integer $i\geq 1$, we denote $\{1,\dots,i\}$ by $[i]$.

In this work, we consider a noisy non-adaptive group testing problem under both combinatorial and probabilistic models. 
In combinatorial model, there are $K$ defective items among a group of $N$ items, whereas in probabilistic model, each item is defective with probability $K/N$, independently from other items. For both models, the parameter $K$ is assumed to be known. %the expected number of defective items is $K$. 
The problem is to identify all %or a sufficiently large fraction of the 
defective items by testing groups of items, with the minimum possible number of tests. The outcome of each test is a binary number. 
The focus of this work is when $N$ is limited rather than on the asymptotic regime.

We define the support vector $\mathbf{x}\in \{0,1\}^N$ to represent the set of $N$ items. The $i$-th component of $\mathbf{x}$, i.e., $x_i$, is $1$ if and only if the $i$-th item is defective. 
In non-adaptive group testing, designing a testing scheme consisting of $M$ tests is equivalent to the construction of a binary matrix with $M$ rows which is referred to as measurement matrix. 
%{\color{red} Shouldn't we first say that we focus on non-adaptive group testing?} 
We let matrix ${\textbf{A}\in \{0,1\}^{M\times N}}$ denote the measurement matrix. 
If $a_{ti}=1$ in the measurement matrix ${\textbf{A}}$, it means that the $i$-th item is present in the $t$-th test. 
The design of measurement matrices for group testing has been studied extensively \cite{CIT-099}.
Our proposed algorithms are applicable for any measurement matrix; however, evaluation results are presented for Bernoulli designs in this paper.

%In this paper, we consider the Bernoulli design for the measurement matrix. 
%{\color{red} Why Bernoulli design? Is this a new model?} 
%In a Bernoulli design, we have $\mathbb{P}(a_{ti}=1)=p$ and $\mathbb{P}(a_{ti}=0)=1-p$ independently over $i\in [N]$ and $t\in [M]$, for some constant $0<p<1$. 
%That is, each item is included in each test independently at random with some fixed probability $p=\nu/K$ where $\nu$ is a constant and $K$ is the number of defective items. {\color{red} What is the role of this $\nu$? Has it been introduced and discussed in the literature?}
 
 The standard noiseless group testing is formulated component-wise using the Boolean OR operation as ${y_t=\bigvee_{i=1}^{N} a_{ti}x_i}$ where $y_t$ and $\bigvee$ are the $t$th test result and a Boolean OR operation, respectively. In this paper, we consider the widely-adopted binary symmetric noise model where the values $\bigvee_{i=1}^{N} a_{ti}x_i$ are flipped independently at random with a given probability. The $t$th test result in a binary symmetric noise model is given by
\[   
     y_t=\begin{cases}
       \bigvee_{i=1}^{N} a_{ti}x_i &\quad\text{with probability}~ 1-\rho,\\
        1\oplus \bigvee_{i=1}^{N} a_{ti}x_i &\quad \text{with probability} ~\rho,\\
       
     \end{cases}
\] 
where $\oplus$ is the XOR operation. 
Note that this model and the proposed algorithms can be easily extended to include the general binary noise model where the values $\bigvee_{i=1}^{N} a_{ti}x_i$ are flipped from $0$ to $1$ and from $1$ to $0$ with different probabilities.
However, for ease of exposition, we focus only on the binary symmetric noise model. %{\color{red} I agree that the model can be extended. But this does not indicate whether the discussed and proposed decoding algorithms can also be extended to such scenarios.}
We let vector $\mathbf{y}\in\{0,1\}^M$ denote the outcomes of the $M$ tests.
The objective is to minimize the number of tests required to identify the set of defective items while meeting a target success probability, false positive and false negative rates.
These metrics are formally defined in Section~\ref{sec:SR}.

%The objective is to successfully identify the set of defective items with 
%high probability given the test results vector $\mathbf{y}$ using minimum number of tests. {\color{red} what is the source of randomness? the sense of "with high probability" needs to be specified clearly.}

\section{Decoding Algorithms \label{sec:main results}}
\subsection{Belief Propagation}
\begin{figure}
\centering
\resizebox{4.5cm}{4.5cm}{
\begin{tikzpicture}
\def\horzgap{2in}; %Horizontal gap between nodes/levels
\def \gapVN{0.5in}; %vertical gap between nodes
\def \gapCN{0.6in}; %Horizontal gap between nodes

\def\nodewidth{0.25in};
\def\nodewidthA{0.25in};
\def \edgewidth{0.01in};
\def\ext{0.1in};
\def \dotwidth{0.3mm};

\tikzstyle{dots} = [rectangle, draw,line width=0.05mm,  inner sep=0mm, fill=black, minimum height=\dotwidth, minimum width=\dotwidth]
\tikzstyle{check} = [rectangle, draw,line width=0.05mm,  inner sep=0mm, fill=white, minimum height=\nodewidthA, minimum width=\nodewidthA]
\tikzstyle{bit0} = [circle, draw, line width=0.05mm, inner sep=0mm, fill=white, minimum size=\nodewidth]
\tikzstyle{bit1} = [circle, draw, line width=0.05mm, inner sep=0mm, fill=blue, minimum size=\nodewidth]
\tikzstyle{bituncover} = [circle, draw=none, line width=0.05mm, inner sep=0mm, fill=gray, minimum size=\nodewidthA]

\tikzstyle{edgesock} = [circle, inner sep=0mm, minimum size=\edgewidth,draw, fill=white]     

\foreach \vn in {1,2,3,4,5,6}{
 \node[bit0] (vn\vn) at (0,-\vn*\gapVN) {};
}                    

\path (vn1) ++(0,0) node()[scale=0.25, inner sep=0mm] {\Huge{$x_{1}$}};

\path (vn2) ++(0,0) node()[scale=0.25, inner sep=0mm] {\Huge{$x_{2}$}};

\path (vn3) ++(0,0) node()[scale=0.25, inner sep=0mm] {\Huge{$x_{3}$}};

\path (vn4) ++(0,0) node()[scale=0.25, inner sep=0mm] {\Huge{$x_{4}$}};

\path (vn5) ++(0,0) node()[scale=0.25, inner sep=0mm] {\Huge{$x_{5}$}};

\path (vn6) ++(0,0) node()[scale=0.25, inner sep=0mm] {\Huge{$x_{6}$}};

\foreach \cn in {1,...,4}{
\node[check] (cn\cn) at (\horzgap,-\cn*\gapCN-0.3in) {};
}

\draw[line width=0.05mm] (vn1.east)--(cn1.west);
\draw[line width=0.05mm] (vn3.east)--(cn1.west);
\draw[line width=0.05mm] (vn5.east)--(cn1.west);

\draw[line width=0.05mm] (vn2.east)--(cn2.west);
\draw[line width=0.05mm] (vn3.east)--(cn2.west);
\draw[line width=0.05mm] (vn4.east)--(cn2.west);
\draw[line width=0.05mm] (vn6.east)--(cn2.west);

\draw[line width=0.05mm] (vn2.east)--(cn3.west);
\draw[line width=0.05mm] (vn4.east)--(cn3.west);
\draw[line width=0.05mm] (vn5.east)--(cn3.west);

\draw[line width=0.05mm] (vn1.east)--(cn4.west);
\draw[line width=0.05mm] (vn4.east)--(cn4.west);
\draw[line width=0.05mm] (vn5.east)--(cn4.west);
\draw[line width=0.05mm] (vn6.east)--(cn4.west);

\def\moveX {3.2*\nodewidth};
\def\moveXA {2*\nodewidth};

%\path (cn1.east)++(\moveX,0) node ()[scale=0.2] {\Huge{Random $+\mathbf{w_{1}}$}};
\path (cn1)++(0,0) node ()[scale=0.25] {\Huge{$y_1$}};
    \path (cn2)++(0,0) node ()[scale=0.25] {\Huge{$y_2$}};
    \path (cn3)++(0,0) node ()[scale=0.25] {\Huge{$y_3$}};
    \path (cn4)++(0,0) node ()[scale=0.25] {\Huge{$y_4$}};
%\path (cn1.east)++(\moveX+0.25in,0.15in) node ()[scale=0.28] {\Huge{$\rightarrow x_{1_1}+x_{1_2}+\cdots+x_{1_r}$}};
%\path (cn1.east)++(\moveX-0.1in,0.15in) node ()[scale=0.25] {\Huge{$\rightarrow 1$}};

\path (vn3.west)++(-0.5,-0.35in) node ()[scale=0.3] {\Huge{Items}};   
\path (cn2.east)++(0.5,-0.35in) node ()[scale=0.3] {\Huge{Tests}};

\end{tikzpicture}
}
\caption{An example of a factor graph representing a group testing scheme.} \label{fig:Bipartite}%\vspace{-0.125cm}
\end{figure}
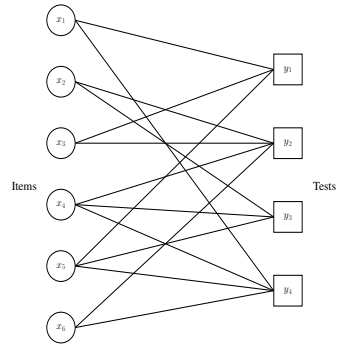

The belief propagation algorithm have gained promising success in different applications in recent years. It has been applied successfully to the problems in the area of coding theory and compressed sensing. Most of these works consider the asymptotic regime; however, we want to apply the belief propagation algorithm to the practical regime where the number of item is limited. To apply the belief propagation algorithm we consider the factor graph (Tanner graph) representation of the group testing scheme as shown in Figure~\ref{fig:Bipartite}. 
In the Tanner graph, there are $N$ nodes at the left side of the graph corresponding to items. Also, there are $M$ nodes at the right side corresponding to the tests. This graph shows the connections between the items and the tests. Each item node is connected to the test nodes that the item participates in, according to the measurement matrix. In general, in a belief propagation algorithm, messages are exchanged between the nodes of the graph. For a loopy belief propagation algorithm, the messages are passed iteratively from items to tests and vice versa. 
We let ${\mu}_{i \rightarrow t} = [\mu_{i\rightarrow t}(0) \ \mu_{i\rightarrow t}(1)]$ and ${\mu}_{t\rightarrow i} = [\mu_{t \rightarrow i}(0) \ \mu_{t\rightarrow i}(1)]$ denote the message from item $i$ to test $t$ and the message from test $t$ to item $i$, respectively, where
%The messages are normalized that such the vector representing the message
%Note that these messages are probability distributions. 
%It has been shown that the messages from items to tests satisfy the following equations:
\begin{equation}\label{eq:itm-to-tst}
\begin{cases}
 \mu_{i\rightarrow t}(0) \propto (1-\frac{K}{N}) \displaystyle \prod_{t^{\prime}\in \mathcal{N}(i)\setminus \{t\}}\mu_{t^{\prime}\rightarrow i}(0), \\
 \mu_{i\rightarrow t}(1) \propto \frac{K}{N} \displaystyle \prod_{t^{\prime}\in \mathcal{N}(i)\setminus \{t\}}\mu_{t^{\prime}\rightarrow i}(1),\\
  \end{cases}
\end{equation}
where $\propto$ indicates equality up to a
normalizing constant, and $\mathcal{N}(i)$ denotes the neighbours of the item node $i$. Note that these messages follow a probability distribution, i.e., $\mu_{i\rightarrow t}(0) + \mu_{i\rightarrow t}(1)=1 $. For both combinatorial and probabilistic models, %Assuming that there are $K$ defective items among a total of $N$ items, 
we initialize the messages by
\begin{equation}\label{eq:intl}
 \mu_{i\rightarrow t}(1)=1-\mu_{i\rightarrow t}(0)=\frac{K}{N}.   
\end{equation}

% {\color{red} (what about the cases in which $K$ is unknown?)}

% {\color{red} What does ``random bipartite graph'' mean here? You can say ``An example of a factor graph representing a group testing scheme.''}

The messages from tests to items are given as follows. If $y_t=0$, we have %{\color{red} what is superscript $r+1$?}
\begin{equation}\label{eq:tst-to-itm0}
\begin{cases}
 \mu_{t\rightarrow i}(0) \propto  \rho + (1-2\rho) \displaystyle \prod_{i^{\prime}\in \mathcal{N}(t)\setminus \{t\}}\mu_{i^{\prime}\rightarrow t}(0), \\
 \mu_{t\rightarrow i}(1) \propto  \rho, \\
  \end{cases}
\end{equation} and if $y_t=1$, we have
\begin{equation}\label{eq:tst-to-itm1}
\begin{cases}
 \mu_{t\rightarrow i}(0) \propto  1-\rho - (1-2\rho) \displaystyle \prod_{i^{\prime}\in \mathcal{N}(t)\setminus \{t\}}\mu_{i^{\prime}\rightarrow t}(0), \\
 \mu_{t\rightarrow i}(1) \propto  1-\rho. \\
  \end{cases}
\end{equation}
Again, these messages follow a probability distribution and it holds that $\mu_{t\rightarrow i}(0) + \mu_{t\rightarrow i}(1)=1 $. In the end, we compute the marginals of the posterior distribution as follows.
\begin{equation}
\begin{cases}
{{q}}(x_{i}=0) \propto  (1-\frac{K}{N}) \displaystyle \prod_{t^{\prime}\in \mathcal{N}(i)}\mu_{t^{\prime}\rightarrow i}(0), \\
  {{q}}(x_{i}=1) \propto  \frac{K}{N} \displaystyle \prod_{t^{\prime}\in \mathcal{N}(i)}\mu_{t^{\prime}\rightarrow i}(1). \\
  \end{cases}
\end{equation}

It is more convenient to compute the Log-Likelihood Ratio (LLR) of a marginal and work with it instead.
\begin{equation}
\lambda_i=\ln \frac{ {{q}}(x_{i}=1)}{{{q}}(x_{i}=0)}=\ln \frac{K}{N-K} + \sum_{t^{\prime}\in \mathcal{N}(i)}\ln \frac{\mu_{t^{\prime}\rightarrow i}(1)}{\mu_{t^{\prime}\rightarrow i}(0)}
\end{equation}

For the combinatorial model, we sort the LLRs of the marginals in decreasing
order and announce the items corresponding to the top $K$ LLRs to be the defective items. For the probabilistic model, we consider a threshold, $\tau$, and items for which $ \lambda_i \geq \tau$ are declared defective. A natural threshold one can choose is $\tau=0$. Note that values other than $0$ are also permissible. Different thresholds result in different false positive and false negative rates. Hence, a suitable choice for the threshold is vital especially for the scenarios in which the false positive rate and false negative rate are not equally important. Algorithm~\ref{alg:BP} defines the belief propagation algorithm.

%\textcolor{red}{KRN comment - you may want to emphasize that in each iteration we do the computation $\forall t \in [T], \forall i \in [N]$}

\begin{algorithm}
\caption{Belief Propagation}
\label{alg:BP}
\label{CHalgorithm}
\begin{algorithmic}[1]
\State Initialize $\mu_{i\rightarrow t}(1)=1-\mu_{i\rightarrow t}(0)=\frac{K}{N}$ $\forall i\in [N], \forall t\in \mathcal{N}(i)$
\For{$j=1,2,\cdots,\text{iter}$ }
\State Compute $\mu_{t\rightarrow i}(0)$ and $\mu_{t\rightarrow i}(1)$ $\forall t\in [M], \forall i\in \mathcal{N}(t)$ %{\color{red} $x_i$'s are not known by the decoder. You may want to say: Compute $\mu_{t\rightarrow i}(0)$ and $\mu_{t\rightarrow i}(1)$ $\forall t\in [M], i\in [N]$. Please make this change in all algorithms.}
\State Compute $\mu_{i\rightarrow t}(0)$ and $\mu_{i\rightarrow t}(1)$ $\forall i\in [N], \forall t\in \mathcal{N}(i)$
\EndFor
\State Compute $\lambda_i$ $\forall  i\in [N]$
\end{algorithmic}
\end{algorithm}

\subsection{Random Scheduling Belief Propagation }
Traditional belief propagation algorithm utilizes flooding scheduling, i.e., in each iteration all messages are updated simultaneously. However, in random scheduling, we update the messages from test nodes to item nodes in a randomized fashion. We start from initialization of the messages from test nodes to their neighboring item nodes as follows. %{\color{red} Why this initialization?}
\begin{equation} \label{eq:intl1}
\mu_{t\rightarrow i}(0)=\mu_{t\rightarrow i}(1)=\frac{1}{2}.
\end{equation}
Also, we initialize the messages from item nodes to their neighboring test nodes as in~\eqref{eq:intl}. 
%Then, we apply~\eqref{eq:itm-to-tst} once to also initialize the messages from item nodes to test nodes. 
After that, for each iteration $j\geq 1$, when we want to update the messages from tests to items, we randomly choose a test node and only send messages from that test node to its neighbouring item nodes. Next, we update the messages from the neighbouring item nodes of this test node. In the end, we identify the defective items in a similar way that was explained for the traditional belief propagation algorithm. Random scheduling belief propagation is formally described in Algorithm~\ref{alg:BPRS}.

\begin{algorithm}
\caption{Random Scheduling Belief Propagation}
\label{alg:BPRS}
\label{CHalgorithm}
\begin{algorithmic}[1]
\State Initialize $\mu_{t\rightarrow i}(0)=\mu_{t\rightarrow i}(1)=\frac{1}{2}$ $\forall t\in [M], \forall i\in \mathcal{N}(t)$ 
\State Initialize $\mu_{i\rightarrow t}(1)=1-\mu_{i\rightarrow t}(0)=\frac{K}{N}$ $\forall i\in [N], \forall t\in \mathcal{N}(i)$ 
\For{$j=1,2,\cdots,\text{iter}$ }
\State Select a test node ${t}^{\prime}$ at random
\State Compute $\mu_{{t}^{\prime}\rightarrow i}(0)$ and $\mu_{{t}^{\prime}\rightarrow i}(1)$ $\forall i\in \mathcal{N}({t}^{\prime})$
\For{each $i\in \mathcal{N}({t}^{\prime})$}
\State Compute $\mu_{i\rightarrow s}(0)$ and $\mu_{i\rightarrow s}(1)$ $\forall s \in \mathcal{N}(i)$
\EndFor
\EndFor
\State Compute $\lambda_i$ $\forall i\in [N]$
\end{algorithmic}
\end{algorithm}

\begin{algorithm}
\caption{Node-wise Residual Belief Propagation}
\label{alg:NWRBP}
\label{CHalgorithm}
\begin{algorithmic}[1]
\State Initialize $\mu_{t\rightarrow i}(0)=\mu_{t\rightarrow i}(1)=\frac{1}{2}$ $\forall t\in [M], \forall i\in \mathcal{N}(t)$ 
\State Initialize $\mu_{i\rightarrow t}(1)=1-\mu_{i\rightarrow t}(0)=\frac{K}{N}$ $\forall i\in [N], \forall t\in \mathcal{N}(i)$
\State Compute $r_{{t}\rightarrow i}$ $\forall t\in [M], \forall i\in \mathcal{N}(t)$ % and generate $Q$ 
\For{$j=1,2,\cdots,\text{iter}$ }
\State Let $\displaystyle t' = \argmax_{t\in [M]} \max_{i\in \mathcal{N}(t)} r_{t\rightarrow i}$ %$r_{{t^{\prime}}\rightarrow i}$ be the first message in $Q$
\For{each $i\in \mathcal{N}({t^{\prime}})$ }
\State Compute $\mu_{t^{\prime}\rightarrow i}(0)$ and $\mu_{t^{\prime}\rightarrow i}(1)$
\State Set $r_{{t^{\prime}}\rightarrow i}=0$ % and reorder $Q$
\For{each $t^{\prime\prime}\in \mathcal{N}(i)\setminus \{t^{\prime}\}$ }
\State Compute $\mu_{i\rightarrow t^{\prime\prime}}(0)$ and $\mu_{i\rightarrow t^{\prime\prime}}(1)$
\For{each $i^{\prime}\in \mathcal{N}(t^{\prime\prime})\setminus \{i\}$ }
\State Compute $r_{{t^{\prime\prime}}\rightarrow i^\prime}$ % and reorder $Q$
\EndFor
\EndFor
\EndFor
\EndFor
\State Compute $\lambda_i$ $\forall i\in [N]$
\end{algorithmic}
\end{algorithm}

\begin{table*}[]
\centering
\begin{tabular}{|c|c|c|c|c|c|c|c|c|c|c|c|c|c|c|}
\hline
\multicolumn{15}{|c|}{$N=100$} \\ \hline
\multirow{2}{*}{$K$} & \multirow{2}{*}{$\rho$} & \multirow{2}{*}{$M$} & \multicolumn{3}{|c|}{BP} & \multicolumn{3}{c|}{RSBP} & \multicolumn{3}{c|}{NW-RBP} & \multicolumn{3}{c|}{Optimal}\\ \cline{4-15} 
 &  &   & Suc. Pr. & FNR   & FPR   &   Suc. Pr.  &   FNR    &   FPR    &   Suc. Pr.    &   FNR    &   FPR    &   Suc. Pr.    &    FNR  & FPR\\ \hline
\multirow{9}{*}{2} & \multirow{3}{*}{0.01} & 20 & 0.8135 & 0.1016 & 0.0021 & 0.8559 & 0.0805 & 0.0016 & 0.8617 & 0.0756 & 0.0016 & 0.8768 & 0.0672 & 0.0014\\ \cline{3-15} 
                   &                       & 25 & 0.9368 & 0.0343 & 0.0007 & 0.9521 & 0.0257 & 0.0005 & 0.9602 & 0.0233 & 0.0005 & 0.9689 & 0.0161 & 0.0003\\ \cline{3-15} 
                   &                       & 30 & 0.9788 & 0.0113 & 0.0002 & 0.9821 & 0.0101 & 0.0002 & 0.9880 & 0.0072 & 0.0001 & 0.9914 & 0.0043 & 0.0001\\ \cline{2-15} 
                   & \multirow{3}{*}{0.03} & 25 & 0.8437 & 0.0874 & 0.0018 & 0.8832 & 0.0659 & 0.0013 & 0.9120 & 0.0480 & 0.0010 & 0.9147 & 0.0463 & 0.0009\\ \cline{3-15} 
                   &                       & 30 & 0.9270 & 0.0389 & 0.0008 & 0.9470 & 0.0298 & 0.0006 & 0.9607 & 0.0211 & 0.0004 & 0.9673 & 0.0171 & 0.0003\\ \cline{3-15} 
                   &                       & 35 & 0.9635 & 0.0197 & 0.0004 & 0.9750 & 0.0140 & 0.0003 & 0.9788 & 0.0113 & 0.0003 & 0.9892 & 0.0054 & 0.0001\\ \cline{2-15} 
                   & \multirow{3}{*}{0.05} & 30 & 0.8580 & 0.0781 & 0.0016 & 0.8893 & 0.0631 & 0.0013 & 0.9216 & 0.0441 & 0.0009 & 0.9325 & 0.0372 & 0.0008\\ \cline{3-15} 
                   &                       & 35 & 0.9151 & 0.0469 & 0.0010 & 0.9441 & 0.0307 & 0.0006 & 0.9574 & 0.0239 & 0.0005 & 0.9701 & 0.0157 & 0.0003\\ \cline{3-15} 
                   &                       & 40 & 0.9524 & 0.0262 & 0.0005 & 0.9620 & 0.0213 & 0.0004 & 0.9818 & 0.0092 & 0.0002 & 0.9854 & 0.0076 & 0.0002\\ \hline
\multirow{9}{*}{4} & \multirow{3}{*}{0.01} & 40 & 0.8037 & 0.0536 & 0.0022 & 0.8395 & 0.0412 & 0.0015 & 0.8359 & 0.0415 & 0.0017 & 0.8388 & 0.0412 & 0.0016\\ \cline{3-15} 
                   &                       & 45 & 0.8961 & 0.0275 & 0.0011 & 0.9040 & 0.0249 & 0.0010 & 0.9055 & 0.0233 & 0.0010 & 0.9145 & 0.0220 & 0.0009\\ \cline{3-15} 
                   &                       & 50 & 0.9465 & 0.0138 & 0.0006 & 0.9471 & 0.0137 & 0.0006 & 0.9482 & 0.0135 & 0.0005 & 0.9488 & 0.0131 & 0.0005\\ \cline{2-15} 
                   & \multirow{3}{*}{0.03} & 50 & 0.8583 & 0.0401 & 0.0017 & 0.8650 & 0.0360 & 0.0015 & 0.8799 & 0.0331 & 0.0014 & 0.8851 & 0.0316 & 0.0013\\ \cline{3-15} 
                   &                       & 55 & 0.9065 & 0.0265 & 0.0011 & 0.9268 & 0.0193 & 0.0009 & 0.9291 & 0.0181 & 0.0008 & 0.9366 & 0.0169 & 0.0007\\ \cline{3-15} 
                   &                       & 60 & 0.9458 & 0.0147 & 0.0006 & 0.9538 & 0.0099 & 0.0004 & 0.9559 & 0.0098 & 0.0004 & 0.9598 & 0.0095 & 0.0004\\ \cline{2-15} 
                   & \multirow{3}{*}{0.05} & 60 & 0.8751 & 0.0365 & 0.0015 & 0.8991 & 0.0294 & 0.0012 & 0.9013 & 0.0283 & 0.0011 & 0.9071 & 0.0242 & 0.0010\\ \cline{3-15} 
                   &                       & 65 & 0.9207 & 0.0225 & 0.0009 & 0.9422 & 0.0165 & 0.0006 & 0.9450 & 0.0146 & 0.0006 & 0.9474 & 0.0136 & 0.0005\\ \cline{3-15} 
                   &                       & 70 & 0.9471 & 0.0191 & 0.0005 & 0.9564 & 0.0116 & 0.0005 & 0.9583 & 0.0109 & 0.0004 & 0.9596 & 0.0104 & 0.0004\\ \hline

\end{tabular}
\caption{{The performance of different decoding algorithms for the combinatorial model when $N=100$.}}
\label{Table:comb1}
\end{table*}

\begin{table*}[]
\centering
\begin{tabular}{|c|c|c|c|c|c|c|c|c|}
\hline
\multicolumn{9}{|c|}{$N=200$} \\ \hline
\multirow{2}{*}{$K$} & \multirow{2}{*}{$\rho$} & \multirow{2}{*}{$M$} & \multicolumn{3}{|c|}{BP} & \multicolumn{3}{c|}{RSBP} \\ \cline{4-9} 
 &  &   & Succ. Prob. & FNR   & FPR   &   Suc. Prob.  &   FNR    &   FPR    \\ \hline

\multirow{9}{*}{4} & \multirow{3}{*}{0.01} & 50 & 0.8762 & 0.0361 & 0.0007 & 0.9163 & 0.0221 & 0.0004 \\ \cline{3-9} 
                   &                       & 55 & 0.9354 & 0.0182 & 0.0004 & 0.9521 & 0.0123 & 0.0002 \\ \cline{3-9} 
                   &                       & 60 & 0.9667 & 0.0098 & 0.0002 & 0.9732 & 0.0069 & 0.0001 \\ \cline{2-9} 
                   & \multirow{3}{*}{0.03} & 60 & 0.8832 & 0.0374 & 0.0008 & 0.9203 & 0.0213 & 0.0004 \\ \cline{3-9} 
                   &                       & 65 & 0.9286 & 0.0228 & 0.0005 & 0.9532 & 0.0121 & 0.0002 \\ \cline{3-9} 
                   &                       & 70 & 0.9526 & 0.0160 & 0.0003 & 0.9771 & 0.0063 & 0.0001 \\ \cline{2-9} 
                   & \multirow{3}{*}{0.05} & 70 & 0.8835 & 0.0405 & 0.0008 & 0.9312 & 0.0189 & 0.0004 \\ \cline{3-9} 
                   &                       & 75 & 0.9252 & 0.0260 & 0.0005 & 0.9348 & 0.0168 & 0.0002 \\ \cline{3-9} 
                   &                       & 80 & 0.9544 & 0.0159 & 0.0003 & 0.9752 & 0.0071 & 0.0001 \\ \hline

\end{tabular}
\caption{{The performance of different decoding algorithms for the combinatorial model when $N=200$.}}
\label{Table:comb2}
\end{table*}

\subsection{Node-wise Residual Belief Propagation}
In order to obtain a better performance over the traditional belief propagation algorithm for decoding LDPC codes, the authors of~\cite{4288829} propose node-wise Residual Belief Propagation (RBP). Here, we adopt this algorithm for the noisy group testing problem. Similar to the random scheduling, in node-wise RBP, we choose a test node in each iteration and only update the messages from that test node to its neighboring item nodes. However, unlike the random scheduling, this test node is not chosen at random. We choose this test node using an ordering metric called the residual. In order to compute the residual for the messages from the test node $t\in [M]$ to the item node $i\in \mathcal{N}(t)$, we first compute $\lambda_{{t}\rightarrow i}=\ln \frac{\mu_{t\rightarrow i}(1)}{\mu_{t\rightarrow i}(0)}$
%\begin{equation}
%\end{equation} 
as the LLR of the most updated messages $\mu_{t\rightarrow i} (0)$ and $\mu_{t\rightarrow i} (1)$ from the test node $t$ to the item node $i$ in previous iterations. 
For each $t\in [M]$ and $i\in \mathcal{N}(t)$, we also compute  $\lambda^{*}_{t\rightarrow i} = \ln\frac{\mu^{*}_{t\rightarrow i}(1)}{\mu^{*}_{t\rightarrow i}(0)}$, %which we call the pseudo-updated LLR of the messages from test node $t$ to item node $i$, 
where $\mu^{*}_{t\rightarrow i}(0)$ and $\mu^{*}_{t\rightarrow i}(1)$ are computed using~\eqref{eq:tst-to-itm0} or~\eqref{eq:tst-to-itm1} depending on $y_t=0$ or $y_t=1$, as the pseudo-updated messages from the test node $t$ to the item node $i$ in the current iteration assuming that the test node $t$ is to be scheduled in this iteration. 
The residual $r_{t\rightarrow i}$ is then defined as the absolute value of the difference between the LLR of the most updated messages in previous iterations and the LLR of the pseudo-updated messages from the test node $t$ to item node $i$, i.e., $r_{{t}\rightarrow i}=\abs {\lambda^{*}_{{t}\rightarrow i}-\lambda_{{t}\rightarrow i}}$.  %{\color{red} It is not clear what an ``update'' means here? How is $\lambda^{*}$ computed exactly?}
%\begin{equation}
%\end{equation}
%where $\lambda^{*}_{{t}\rightarrow i}$ denotes the LLR after an update. 
%In each iteration, we form a queue, $Q$, that contains all the residuals in a descending order.

In each iteration,
we select a test node $t$ such that $r_{t\rightarrow i}$ has the highest value among all $t\in [M]$ and all $i\in \mathcal{N}(t)$, and update the messages from the selected test node to its neighboring item nodes. 
(Note that the messages from other test nodes to their neighboring item nodes will not be updated in this iteration.) 
%The corresponding to the highest residual and update the messages of that test node. 
%{\color{red} Residuals are defined for messages. What does "test node corresponding to the highest residual" mean exactly?} 
The idea behind this strategy is that the differences between the LLRs before and after an update approaches zero as loopy BP converges. 
Hence, a large residual means that the corresponding test node is located in a part of the graph that has not converged yet~\cite{4288829}. %{\color{red} The last two sentences seem to be borrowed from Wesel's paper. Please rephrase and cite.} 
%In each iteration, we form a queue, $Q$, that contains all the residuals in a descending order. % ordered by the values of the residuals. {\color{red} What ordering is used? Please clarify.} 
%In each iteration, the test node corresponding to the first residual in $Q$ is selected and its messages to the neighbouring node will be updated. Then, $Q$ is reordered using the new information. {\color{red} This statement is not clear. What is ``new information''? Please explain clearly how ``re-ordering'' is performed.} 
Node-wise RBP is formally described in Algorithm~\ref{alg:NWRBP}.

\section{Simulation Results}\label{sec:SR}
In this section, we compare the performance of the BP algorithm, the Random Scheduling BP (RSBP) algorithm, and the Node-Wise Residual BP (NW-RBP) algorithm for both combinatorial and probabilistic models of the defective items using three metrics. %{\color{red} Please define the terms RSBP and NW-RBP.} 
The first metric, success probability, shows the probability that an algorithm identifies all the defective items correctly. We also use False-Negative Rate (FNR) and False-Positive Rate (FPR) in our performance comparison. 
FNR is defined as the ratio of the number of defective items falsely classified as non-defective over the number of all defective items. 
Similarly, FPR is defined as the number of non-defective items falsely classified as defective over the number of all non-defective items. 
In our simulations, we consider the binary symmetric noise model with parameter $\rho$. %{\color{red} Why this value of $\nu$?} 
Each result is averaged over $3000$ experiments.

The measurement matrix is constructed according to a Bernoulli design \cite{CIT-099}.
In a Bernoulli design, %we have $\mathbb{P}(a_{ti}=1)=\nu/K$ and $\mathbb{P}(a_{ti}=0)=1-\nu/K$ independently over $i\in [N]$ and $t\in [M]$, for some constant $\nu$.  
%That is, 
each item is included in each test independently at random with some fixed probability $\nu/K$ where $K$ is the number of defective items and we set $\nu=\ln 2$. %{\color{red} What is the role of this $\nu$? Has it been introduced and discussed in the literature?}

\begin{table*}[]
\centering
\begin{tabular}{|c|c|c|c|c|c|c|c|c|c|c|c|c|c|c|}
\hline
\multicolumn{15}{|c|}{$N=100$} \\ \hline
\multirow{2}{*}{$K$} & \multirow{2}{*}{$\rho$} & \multirow{2}{*}{$M$} & \multicolumn{3}{|c|}{BP} & \multicolumn{3}{c|}{RSBP} & \multicolumn{3}{c|}{NW-RBP} & \multicolumn{3}{c|}{Optimal}\\ \cline{4-15} 
 &  &   & Suc. Pr. & FNR   & FPR   &   Suc. Pr.  &   FNR    &   FPR    &   Suc. Pr.    &   FNR    &   FPR    &   Suc. Pr.    &    FNR  & FPR\\ \hline
\multirow{9}{*}{2} & \multirow{3}{*}{0.01} & 20 & 0.6282 & 0.2747 & 0.0017 & 0.8033 & 0.0966 & 0.0016 & 0.8150 & 0.0857 & 0.0016 & 0.8248 & 0.0789 & 0.0018\\ \cline{3-15} 
                   &                       & 25 & 0.8041 & 0.1432 & 0.0010 & 0.9207 & 0.0277 & 0.0009 & 0.9238 & 0.0258 & 0.0008 & 0.9296 & 0.0215 & 0.0007\\ \cline{3-15} 
                   &                       & 30 & 0.9035 & 0.0706 & 0.0004 & 0.9568 & 0.0111 & 0.0004 & 0.9669 & 0.0092 & 0.0004 & 0.9711 & 0.0053 & 0.0003\\ \cline{2-15} 
                   & \multirow{3}{*}{0.03} & 25 & 0.6621 & 0.2465 & 0.0020 & 0.8399 & 0.0867 & 0.0012 & 0.8709 & 0.0635 & 0.0007 & 0.8822 & 0.0603 & 0.0008\\ \cline{3-15} 
                   &                       & 30 & 0.7982 & 0.1573 & 0.0008 & 0.9174 & 0.0347 & 0.0007 & 0.9411 & 0.0250 & 0.0004 & 0.9464 & 0.0246 & 0.0004\\ \cline{3-15} 
                   &                       & 35 & 0.8812 & 0.0967 & 0.0003 & 0.9486 & 0.0169 & 0.0005 & 0.9655 & 0.0112 & 0.0003 & 0.9704 & 0.0092 & 0.0002\\ \cline{2-15} 
                   & \multirow{3}{*}{0.05} & 30 & 0.6913 & 0.2349 & 0.0018 & 0.8346 & 0.0802 & 0.0012 & 0.8810 & 0.0554 & 0.0009 & 0.9007 & 0.0457 & 0.0009\\ \cline{3-15} 
                   &                       & 35 & 0.7931 & 0.1613 & 0.0009 & 0.9007 & 0.0396 & 0.0008 & 0.9268 & 0.0305 & 0.0006 & 0.9329 & 0.0256 & 0.0005\\ \cline{3-15} 
                   &                       & 40 & 0.8665 & 0.1059 & 0.0005 & 0.9291 & 0.0224 & 0.0005 & 0.9669 & 0.0115 & 0.0002 & 0.9714 & 0.0091 & 0.0002\\ \hline
                   
%\multirow{9}{*}{4} & \multirow{3}{*}{0.01} & 40 & 0.6856 & 0.0988 & 0.0014 & 0.7460 & 0.0592 & 0.0014 & 0.7700 & 0.0450 & 0.0015 &  0.7573     & 0.0508       & 0.0022\\ \cline{3-15} 
                   %&                       & 45 & 0.7908 & 0.0541 & 0.0011 & 0.8140 & 0.0340 & 0.0010 & 0.8200 & 0.0362 & 0.0010 & 0.8377      &   0.0280    & 0.0010\\ \cline{3-15}
                   %&                       & 50 & 0.8616 & 0.0304 & 0.0008 & 0.8830 & 0.0179 & 0.0008 & 0.8912 & 0.0164 & 0.0007 &   0.8918    &   0.0156    & 0.0009\\ \cline{2-15} 
                   %& \multirow{3}{*}{0.03} & 50 & 0.7278 & 0.0689 & 0.0018 & 0.7430 & 0.0425 & 0.0019 & 0.7538 & 0.0338 & 0.0018 &       &       &\\ \cline{3-15} 
                   %&                       & 55 & 0.7835 & 0.0499 & 0.0016 & 0.8210 & 0.0225 & 0.0015 & 0.8160 & 0.2080 & 0.0014 &       &       &\\ \cline{3-15} 
                   %&                       & 60 & 0.8205 & 0.0318 & 0.0013 & 0.8430 & 0.0150 & 0.0013 & 0.8600 & 0.0143 & 0.0013 &       &       &\\ \cline{2-15} 
                   %& \multirow{3}{*}{0.05} & 60 & 0.7430 & 0.0739 & 0.0011 & 0.7730 & 0.0485 & 0.0011 & 0.7914 & 0.0452 & 0.0010  &       &       &\\ \cline{3-15} 
                   %&                       & 65 & 0.8285 & 0.0466 & 0.0008 & 0.8440 & 0.0262 & 0.0007 & 0.8550 & 0.0246 & 0.0006 &       &       &\\ \cline{3-15} 
                   %&                       & 70 & 0.8665 & 0.0339 & 0.0007 & 0.8827 & 0.0239 & 0.0006 & 0.8930 & 0.0198 & 0.0004 &       &       &\\ \hline

\end{tabular}
\caption{{The performance of different decoding algorithms for the probabilistic model when $N=100$.}}
\label{Table:prob1}
\end{table*}

\begin{table*}[]
\centering
\begin{tabular}{|c|c|c|c|c|c|c|c|c|}
\hline
\multicolumn{9}{|c|}{$N=200$} \\ \hline
\multirow{2}{*}{$K$} & \multirow{2}{*}{$\rho$} & \multirow{2}{*}{$M$} & \multicolumn{3}{|c|}{BP} & \multicolumn{3}{c|}{RSBP} \\ \cline{4-9} 
 &  &   & Succ. Prob. & FNR   & FPR   &   Succ. Prob.  &   FNR    &   FPR   \\ \hline

\multirow{9}{*}{4} & \multirow{3}{*}{0.01} & 50 & 0.7712 & 0.0858 & 0.0007 & 0.8456 & 0.0265 & 0.0005 \\ \cline{3-9} 
                   &                       & 55 & 0.8518 & 0.0467 & 0.0004 & 0.8881 & 0.0166 & 0.0004 \\ \cline{3-9} 
                   &                       & 60 & 0.9017 & 0.0255 & 0.0003 & 0.9238 & 0.0082 & 0.0003 \\ \cline{2-9} 
                   & \multirow{3}{*}{0.03} & 60 & 0.7744 & 0.0911 & 0.0007 & 0.8363 & 0.0312 & 0.0005 \\ \cline{3-9} 
                   &                       & 65 & 0.8415 & 0.0619 & 0.0004 & 0.8906 & 0.0193 & 0.0003 \\ \cline{3-9} 
                   &                       & 70 & 0.8882 & 0.0409 & 0.0003 & 0.9189 & 0.0144 & 0.0002 \\ \cline{2-9} 
                   & \multirow{3}{*}{0.05} & 70 & 0.7889 & 0.0835 & 0.0008 & 0.8487 & 0.0291 & 0.0004 \\ \cline{3-9} 
                   &                       & 75 & 0.8512 & 0.0550 & 0.0005 & 0.8819 & 0.0202 & 0.0003 \\ \cline{3-9} 
                   &                       & 80 & 0.8947 & 0.0396 & 0.0003 & 0.9207 & 0.0138 & 0.0002 \\ \hline

\end{tabular}
\caption{{The performance of different decoding algorithms for the probabilistic model when $N=200$.}}
\label{Table:prob2}
\end{table*}

\subsection{Combinatorial Model}
In Table~\ref{Table:comb1}, we present experimental simulation results for $N=100$ items and $K=2,4$ defective items. Corresponding to each value of $K$, we consider three different values for the noise parameter $\rho=0.01,0.03,0.05$. And for each value of $\rho$, we consider three different number of tests. The optimal decoder which is used as a benchmark here is the maximum-likelihood decoder. The results in Table~\ref{Table:comb1} show that the NW-RBP algorithm outperforms the BP algorithm and the RSBP algorithm for all problem parameters being considered, in the sense of success probability, FNR, and FPR. Specifically, the FNR is reduced drastically for the NW-RBP algorithm compared to the BP algorithm. For some set of parameters, the FNR reduces by about $50\%$ or even more. Note that the reduction in FNR decreases when we go from $K=2$ to $K=4$.
The NW-RBP algorithm perform fairly close to the optimal decoder for a large set of parameters, particularly when the success probability approaches one. 
Note that although the RSBP algorithm is inferior to the NW-RBP algorithm, it outperforms the BP algorithm in the sense of success probability, FNR, and FPR greatly. In Table~\ref{Table:comb2}, we compare the performance of the BP algorithm and the RSBP algorithm for $N=200$ items and $K=4$ items. Similar to the Table~\ref{Table:comb1}, we consider three different values for the noise parameter $\rho$, and three different number of tests for each value of the noise parameter. The results reveal that the RSBP algorithm outperforms the BP algorithm significantly for all parameters being considered. %{\color{red} or a fixed prevalence, or a fixed number of defective items, how does the performance improvement of RS-BP or NW-RBP over BP change?}

\subsection{Probabilistic Model}
In Table~\ref{Table:prob1}, we provide experimental simulation results for $N=100$ items and the the expected number of items $K = 2$. Similar to Table~\ref{Table:comb1}, we consider three different values for the noise parameter $\rho$, and for each value of $\rho$, we consider three different number of tests. A threshold value of $\tau=0$ is considered for BP, RSBP, and NW-RBP algorithms. Similar to the combinatorial model, the NW-RBP algorithm outperforms the BP and the RSBP algorithms for all parameters being tested, in the sense of success probability, FNR, and FPR. It should be noted that the improvement that is achieved by using the NW-RBP algorithm is more considerable than that in the combinatorial model. Also, the reduction in FNR when the NW-RBP algorithm is being used is far more remarkable than that in the combinatorial model. There are many examples in Table~\ref{Table:prob1} for which the FNR reduces by about $80\%$ when the NW-RBP algorithm is used instead of the BP algorithm. We compare the performance of the BP algorithm and the RSBP algorithm for $N=200$ items and the expected number of items $K=4$ in Table~\ref{Table:prob2}. It can be easily seen that the RSBP algorithm outperforms the BP algorithm for the all tested parameters. %{\color{red} For the tested parameters, what are the minimum and maximum improvements observed?}

\begin{figure*}[t!]

\begin{subfigure}[t]{.5\textwidth}
  \centering
   \includegraphics[width=.8\linewidth]{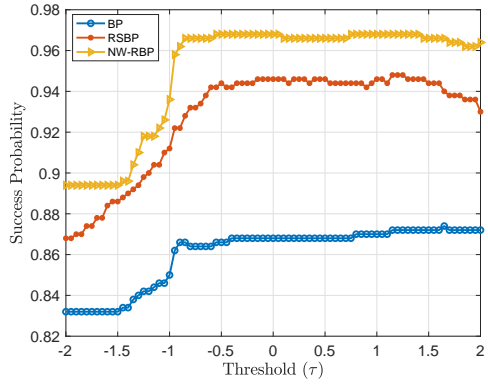}  
  \caption{Success Probability vs. Threshold ($\tau$)}
  \label{fig:sub-first}
\end{subfigure}
\begin{subfigure}[t]{.5\textwidth}
  \centering
 \includegraphics[width=.8\linewidth]{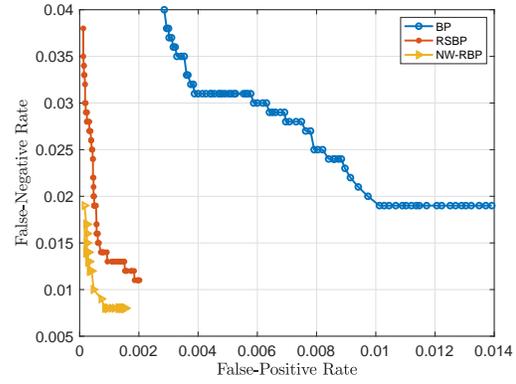}  
  \caption{False-Negative Rate vs. False-Positive Rate}
  \label{fig:sub-second}
\end{subfigure}

\begin{subfigure}[t]{.5\textwidth}
  \centering
  % include third image
  \includegraphics[width=.8\linewidth]{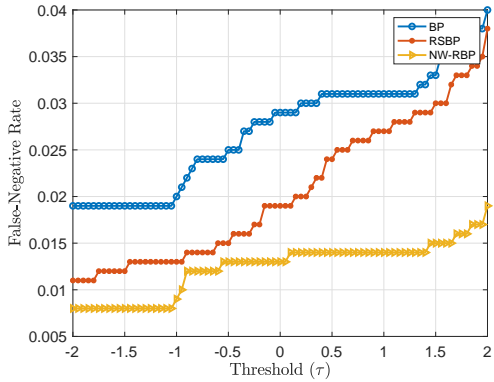}  
  \caption{False-Negative Rate vs. Threshold ($\tau$)}
  \label{fig:sub-third}
\end{subfigure}
\begin{subfigure}[t]{.5\textwidth}
  \centering
  % include fourth image
  \includegraphics[width=.8\linewidth]{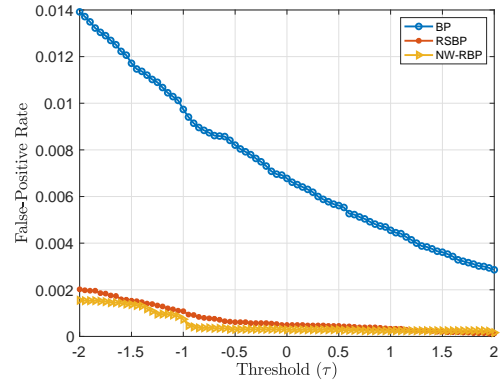}  
  \caption{False-Positive Rate vs. Threshold ($\tau$)}
  \label{fig:sub-fourth}
\end{subfigure}
\caption{The performance of different decoding algorithms for different values of the threshold $\tau$.}
%\caption{Experimental simulations for the symmetric noise model with parameter $\rho = 0.05$, with $N=100$ items, the expected number of items $K = 2$, and $M=40$ tests. }
\label{fig:one}
\label{fig:fig}
\end{figure*}

In Fig.~\ref{fig:one}, we plot experimental simulation results for $N = 100$ items, the expected number of items $K = 2$, and $M=40$ tests, under the symmetric noise model with parameter $\rho = 0.05$. Fig.~\ref{fig:sub-first} depicts success probability as a function of threshold $\tau$. 
It can be observed that the NW-RBP outperforms the BP and RSBP algorithms for all values of $\tau$. However, the difference between the success probabilities of these algorithms varies with $\tau$. It can also be seen that the success probability of the NW-RBP algorithm is at its maximum for $-0.5<\tau<1.5$. 
Fig.~\ref{fig:sub-second} depicts the FNR and FPR of different decoding algorithms for different values of $\tau$. 
That is, 
%The FNR is illustrated as a function of the FPR in Fig.~\ref{fig:sub-second}. 
each point corresponds to the FNR and FPR of a decoding algorithm for the same value of $\tau$. 
The parameter $\tau$ ranges from $-2$ to $2$. 
We have also depicted the FNR and the FPR of these decoding algorithms, as a function of $\tau$ in Fig.~\ref{fig:sub-third} and Fig.~\ref{fig:sub-fourth}, respectively. 
It can be observed that there is a trade-off between the FNR and the FPR. 
Depending on the application at hand, one can choose a threshold value $\tau$ that yields the desired FNR or FPR. For instance, in epidemic detection, the FNR is more important as false-negative errors hugely increase the risk of the spread of infection.

%{\color{red} What are the takeaway messages here? What do we learn from these figures?} {\color{red} Figures do not have caption.}

\bibliographystyle{IEEEtran}
\bibliography{NGTRefs}

\end{document}